\documentclass{PoS}
\usepackage{amsmath,amssymb}

\newcommand{\tmtextbf}[1]{{\bfseries{#1}}}

\newcommand{\bea}{\begin{eqnarray}}
\newcommand{\eea}{\end{eqnarray}}

\newcommand{\fig}[1]{Fig.~\ref{#1}}
\newcommand{\ev}[1]{\langle #1 \rangle}
\newcommand{\re}{{\rm e}}

\newcommand{\rD}{{\rm D}}
\newcommand{\Seff}{S_{\rm eff}}
\newcommand{\oV}{\overline V}
\newcommand{\oH}{\overline H}

\newcommand{\Real}{{\rm Re}\,}
\def\tr{{\rm tr}}

\title{The Lattice Mean-Field Approximation of Gauge-Higgs Unification on the Orbifold}

\ShortTitle{The Lattice Mean-Field Approximation of Gauge-Higgs Unification on the Orbifold}

\author{\speaker{Kyoko Yoneyama} and Francesco Knechtli\\
        Department of Physics, Bergische Universit\"at Wuppertal\\
        Gaussstr. 20 \\ 
        D-42119 Wuppertal, Germany \\
        E-mail: \email{yoneyama@physik.uni-wuppertal.de,knechtli@physik.uni-wuppertal.de}}
        
        \author{Nikos Irges\\
        Department of Physics, National Technical University
        of Athens,
        \\
         GR-157 80 Zografou, Attikis Greece
        \\
        E-mail: \email{irges@mail.ntua.gr}}

\abstract{
A possible extension of the Standard Model of elementary particles is Gauge-Higgs unification, where the Higgs field is identified with (some of) the extra dimensional components of a five-dimensional gauge field. In this scenario there is evidence for the potential and the mass of the Higgs field to be finite. Here we show the behavior of the static potential of a five-dimensional SU(2) lattice gauge theory with orbifold boundary conditions. The potentials are computed within the mean-field approximation including first
order corrections.
\begin{flushright} WUB/11-18 \end{flushright}
}

\FullConference{The XXIX International Symposium on Lattice Field Theory, Lattice2011\\
		July 10-16, 2011\\
		Squaw Valley, Lake Tahoe, California}
\begin{document}
\section{Introduction}
Five-dimensional gauge theory is one of the attractive extensions of the Standard Model. As it is well known, five-dimensional gauge symmetry is expected to keep the Higgs potential finite. And the Higgs potential might cause spontaneous symmetry breaking \cite{Hosotani:1983vg,Hosotani:1989vg}. However, five-dimensional gauge theories are non-renormalizable. Therefore, a finite cut-off is needed. Usually, a finite cut-off cannot preserve gauge invariance. Only on the lattice it is possible to have a gauge invariant finite cut-off.
The five-dimensional $SU(2)$ gauge theory on the torus has been studied using the mean-field method in \cite{Irges:2009bi,Irges:2010qp}. In this work we study the case of orbifold boundary conditions.
\section{Orbifold boundary condition}
The definition of the $S^1/\mathbb{Z}_2$ orbifold for Yang-Mills theories on a five-dimensional Euclidean lattice was constructed in \cite{Irges:2005bi}.
For torus boundary condition, the action has a reflection symmetry and a group conjugation symmetry. The action is invariant under these transformations. 
\bea
{\cal R} : \ && z=(x_{\mu}, x_5) \ \rightarrow \ \bar{z}=(x_{\mu}, -x_5)\\
&& A_{M}(z) \ \rightarrow \ \alpha_M A_M(\bar{z}), \ \alpha_{\mu}=1, \ \alpha_5=-1\\
{\cal C} :  \ && A_M(z) \rightarrow g A_M(z) g^{-1}
\eea
Then, the orbifold projection identifies the fields under the transformation ${\cal R} \cdot {\cal C}$ :
\bea
A_M(z) =  \alpha_M g A_M(\bar{z}) g^{-1}
\eea
 Here we should be careful at the two points $x_5=0, \pi R$, where $R$ is the radius of the fifth dimension. These two points are boundaries of the orbifold. At the boundaries, the field is projected onto itself. Thus only the even components of the gauge field i.e. the components which do not change under ${\cal R} \cdot {\cal C}$ are non-zero. We choose $g=-i\sigma^3$ for gauge group $SU(2)$, where $\sigma^3$ is the Pauli matrix. It follows that only the fields $A_5^1$, $A_5^2$ and $A_{\mu}^3$ are non-zero on the boundaries. The Higgs complex field is constructed with $A_5^1$ and $A_5^2$.

\section{The mean-field expansion} 
The mean-field expansion for gauge theories is reviewed in 
\cite{Drouffe:1983fv}.
$SU(2)$ gauge inks $U$ in the partition function are replaced by $N \times N$ complex matrices $V$ and Lagrange multipliers H :
\bea
\ev{{\cal O}[U]} & = &
\frac{1}{Z} \int \rD V \int \rD H \, {\cal O}[V] \re^{-\Seff[V,H]} \\
\Seff & = & S_G[V] + u(H) + (1/N)\Real\tr\{HV\} \ \\
\re^{-u(H)} & = & \int \rD U \, \re^{(1/N)\Real\tr\{UH\}} \,,
\eea
where $\Real\tr\{UH\} =\sum_n \sum_M \Real\tr\{U(n,M)H(n,M)\}$, $n$ labels the lattice points and M is the five-dimensional direction index. 
The mean-field saddle point (or background) is defined by the minimization of
the classical effective action in terms of constant fields proportional to the identity in group space
\bea
H\longrightarrow \oH\mathbf{1} \,;&
V\longrightarrow \oV\mathbf{1} \,;&
\Seff[\oV,\oH]\;\mbox{=minimal} \,. \label{mfbg}
\eea
Corrections are calculated from Gaussian fluctuations around the saddle point
solution
\bea
H = \oH + h \;& \mbox{and}\; & V = \oV + v \,.
\eea
We impose a covariant gauge fixing on $v$. In \cite{Ruhl:1982er} it was shown
that this is equivalent to gauge-fix the original links $U$.

\begin{figure}[h!]\centering
  \resizebox{7cm}{!}{\includegraphics{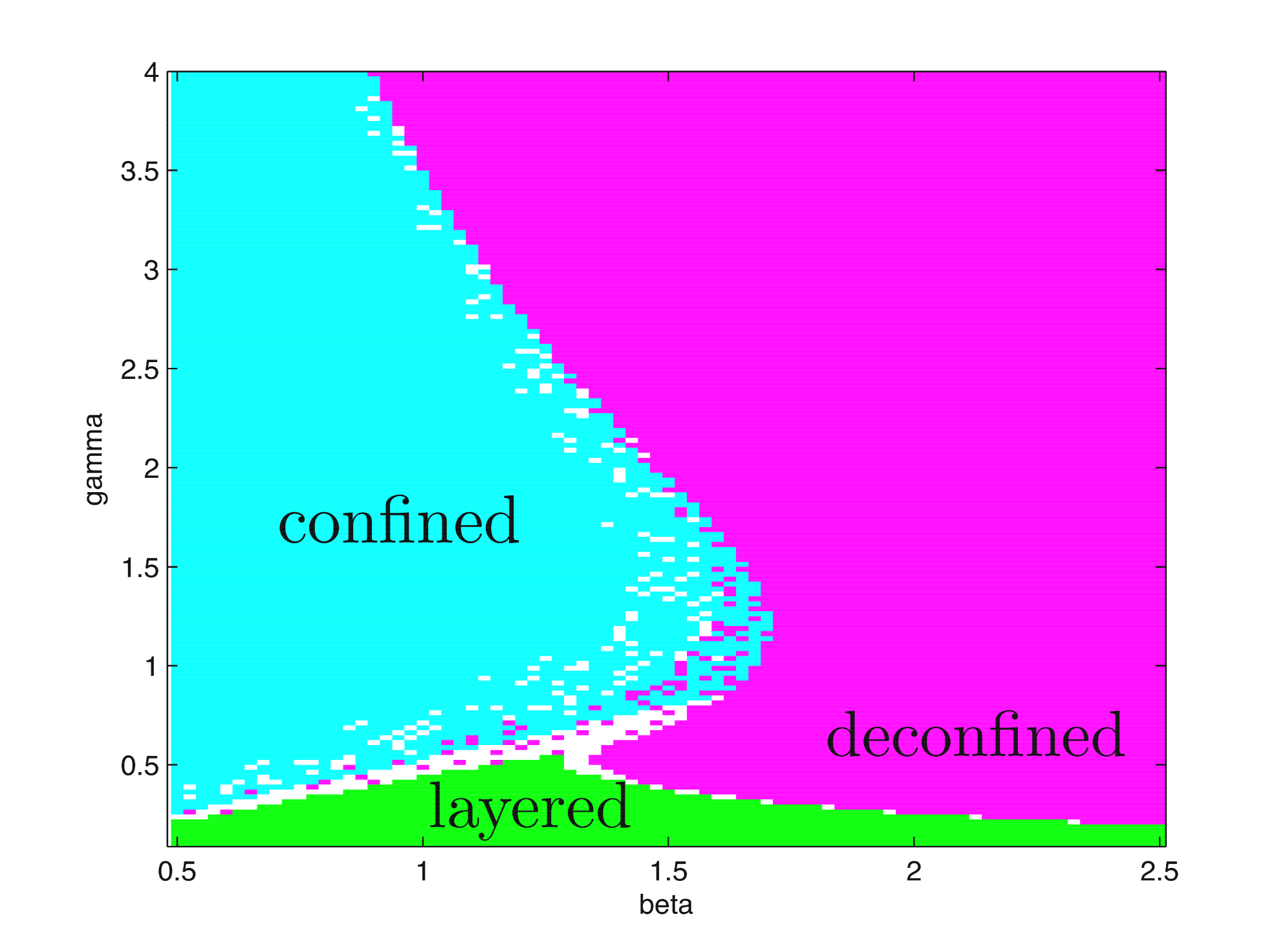}}
  \caption{The mean-field phase diagram for $N_5=8$}
  \label{f_ph}
\end{figure}

Our setup is a
$SU(2)$ gauge theory  formulated on a 
$L_T\times L^3\times N_5$ Euclidean orbifolded lattice
with anisotropic Wilson plaquette action \cite{Wilson:1974sk}
\bea
S_w[U_2,U_1] = S_{w1}[U_2] +S_{w2}[U_1] +S_{w3}[U_2,U_1] \  ; \ \ \  U_1 \in U(1), \ U_2 \in SU(2) .
\eea
Here we define,
\bea
S_{W1}[U_2]&=&\frac{\beta_4}{2N}\sum_{n_\mu}\sum^{N_5-1}_{n_5=1}\sum_{\nu,\rho} {\rm tr} \{ 1-U(n,\nu,\rho) \} +\frac{\beta_5}{N} \sum_{n_\mu}\sum^{N_5-2}_{n_5=1}\sum_{\nu} {\rm tr} \{ 1-U(n,\nu,5) \}\\
S_{W2}[U_1]&=&\frac{\beta_4}{4N}\sum_{n_\mu}\sum_{n_5=0,N_5}\sum_{\nu,\rho} {\rm tr} \{ 1-U(n,\nu,\rho) \}\\
S_{W3}[U_2,U_1]&=&\frac{\beta_5}{N} \sum_{n_\mu}\sum_{n_5=0,N5-1}\sum_{\nu} {\rm tr} \{ 1-U(n,\nu,5) \}.
\eea
We use the reality of the trace in $SU(2)$.
We parametrize the saddle point solution as $\bar{H}(n,\mu)=\bar{h}_0(n_5){\bf 1}, \ \bar{V}(n,\mu)=\bar{v}_0(n_5){\bf 1}$ for $n_5=0,1,\cdots,N_5$ (4d links) and $\bar{H}(n,5)=\bar{h}_0(n_5+1/2){\bf 1}, \ \bar{V}(n,5)=\bar{v}_0(n_5+1/2){\bf 1}$ for $n_5=0,2,\cdots,N_5-1$ (extra dimensional links) \cite{knechtli:2005}.
The phase diagram of the theory [ \fig{f_ph} ] is mapped through
the values of the mean-field 
background, where $\beta=\sqrt{\beta_4 \cdot \beta_5}$ and $\gamma=\sqrt{\beta_5/\beta_4}$. There is a confined phase ($\bar{v}_0(n_5)=0$ for all $n_5$),
a layered phase ($\bar{v}_0(n_5)\neq0$ for $n_5=0,1, \cdots N_5$, $\bar{v}_0(n_5)=0$ for $n_5=1/2,3/2, \cdots N_5-1/2$)
and a deconfined phase ($\bar{v}_0(n_5)\neq0$ for all $n_5$).

\section{Gauge boson mass and static potentials}
We calculate the vector boson mass $m_v$, the scalar mass $m_H$ and the static potential of five-dimensional $SU(2)$ gauge theory on the orbifold. Here we define lines of constant physics by keeping
\bea
\gamma=0.55 \  {\rm and} \ \rho=m_v/m_H =0.65
\eea
constant. And we define a physical scale $r_s$ \cite{Sommer:1993ce} from the static force $F_4$ through the equation $r^2F_4(r) |_{r=r_s}=s=0.2$.

\begin{figure}[h!]\centering
  \resizebox{7cm}{!}{\includegraphics{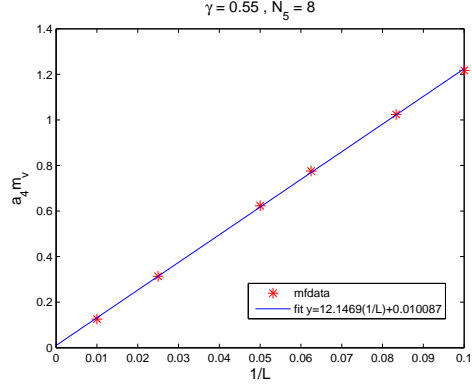}}
  \caption{The vector boson mass.}
  \label{f_vm}
\end{figure}

We construct the gauge boson operator by using the scalar operator which is the projected Polyakov loop. \fig{f_vm} is the plot of the gauge boson mass as the function of $1/L$. The red points are the mean-field data. The blue line is the linear fit of these data. The vector boson mass essentially depends only on $L$ through $a_4m_v\simeq c_L/L$ with $c_L=12.15\simeq 4\pi$. The possible explanation is that this vector boson is the bound state of two gauge bosons with mass $m_w=0$. The bound state mass is $m_v(L)=2\sqrt{m_w^2+\overrightarrow{p}}$ and $m_v=4\pi/L$ for $m_w=0$ and $\overrightarrow{p}=(2\pi/L,0,0)$ 1st nonzero lattice momentum. For $L$ equal infinity, the gauge boson mass is zero or a very small value. It means there is no spontaneous symmetry breaking. 

\begin{figure}[h!]\centering
  \resizebox{7cm}{!}{\includegraphics{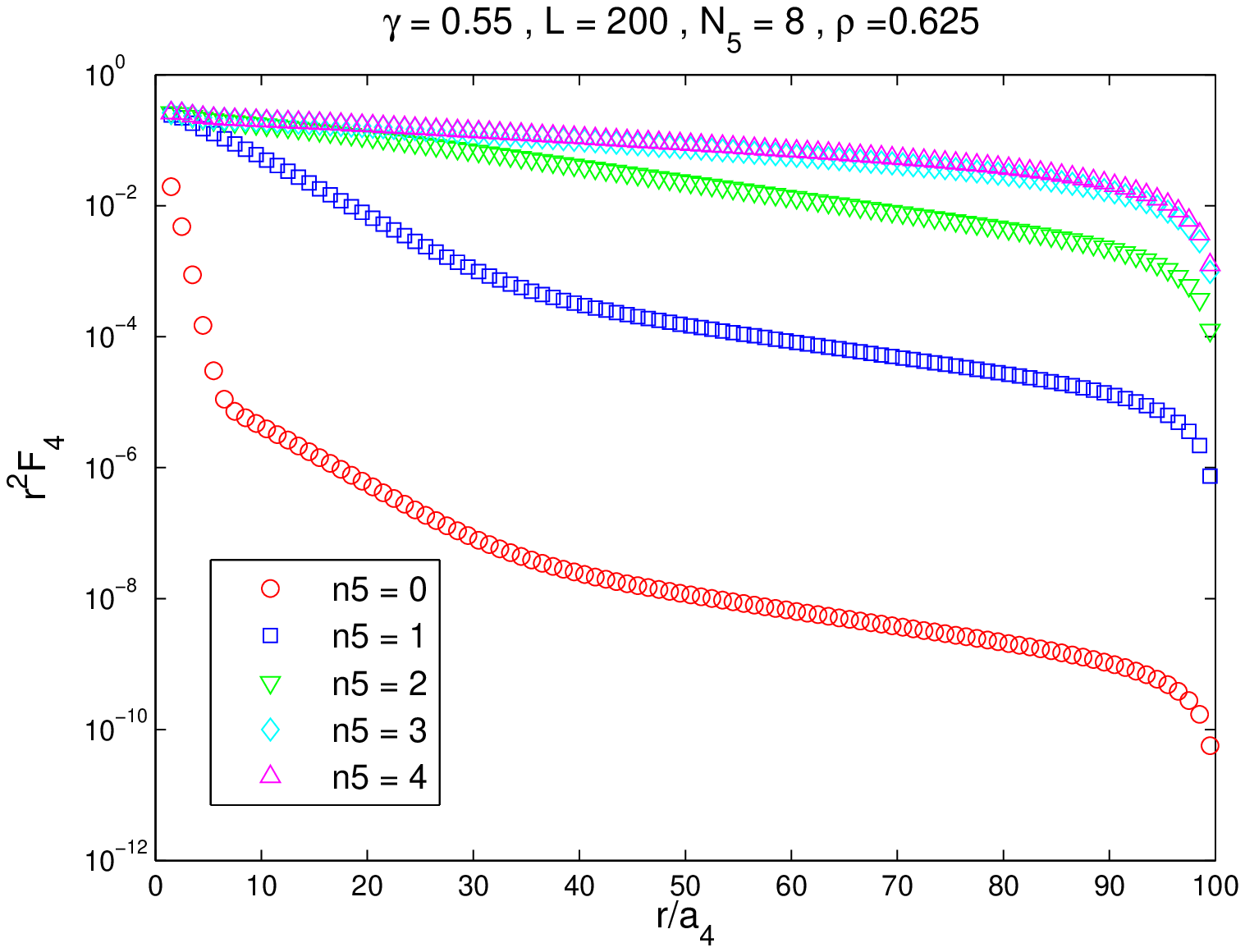}} \ \
  \resizebox{7cm}{!}{\includegraphics{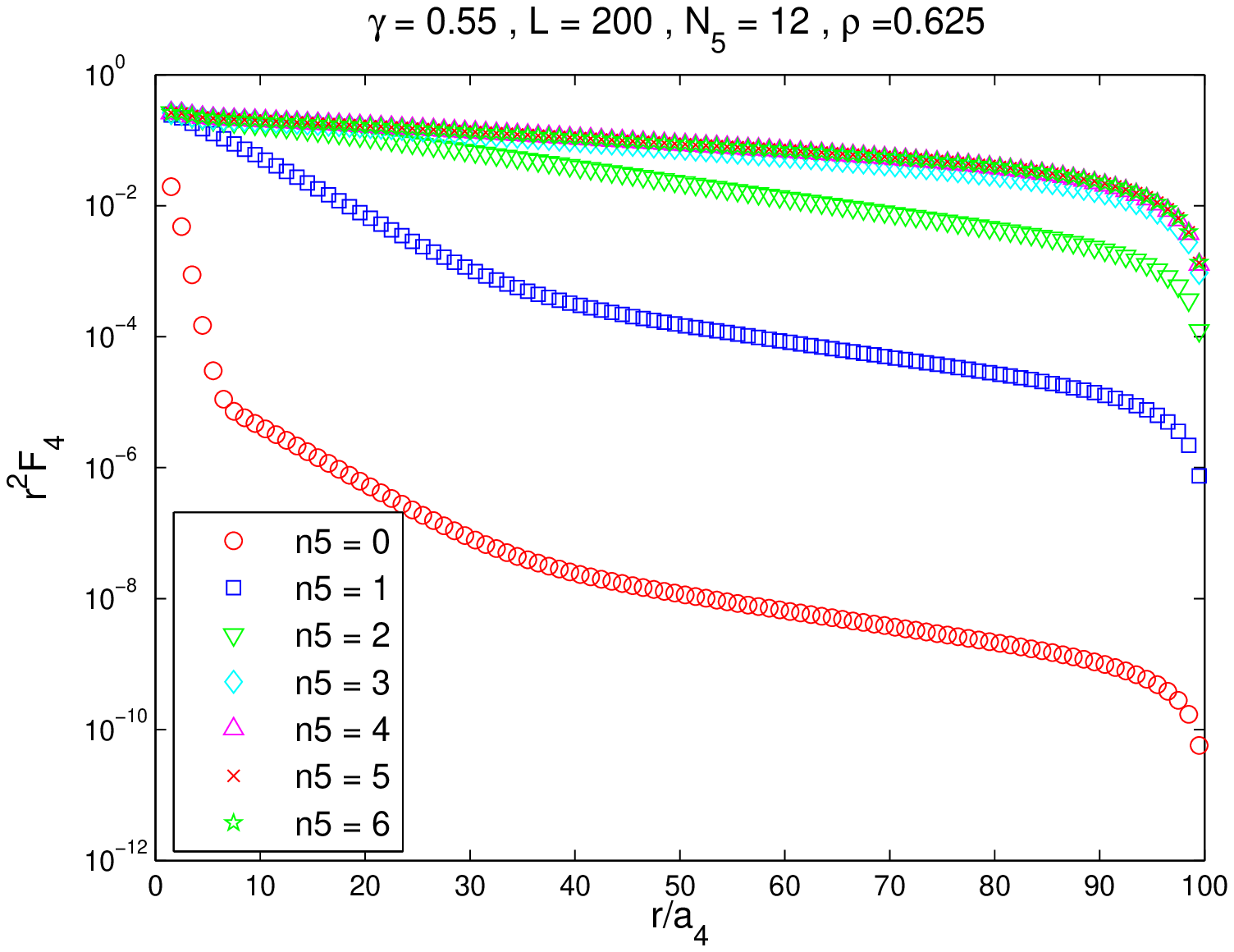}}
  \caption{The four-dimensional static force in each four-dimensional hyperplane for $N_5=8$ and 12.
} 
\label{f_4p}
\end{figure}

\fig{f_4p} shows the static force $F_4=(V_4(r+a_4)-V_4(r))/a_4$, where $V_4$ is the potential along four-dimensional hyperplanes. $n_5$ is the coordinate of the hyperplane along the fifth dimension. The curves are $r^2F_4$ on each four-dimensional slice from the boundary to the middle of the bulk. For the two values of $N_5=8, 12$, it is very similar at the slices $n_5=0,1,2$. They have different behavior from the slice $n5\geq 3$.

\begin{figure}[h!]\centering
  \resizebox{7cm}{!}{\includegraphics{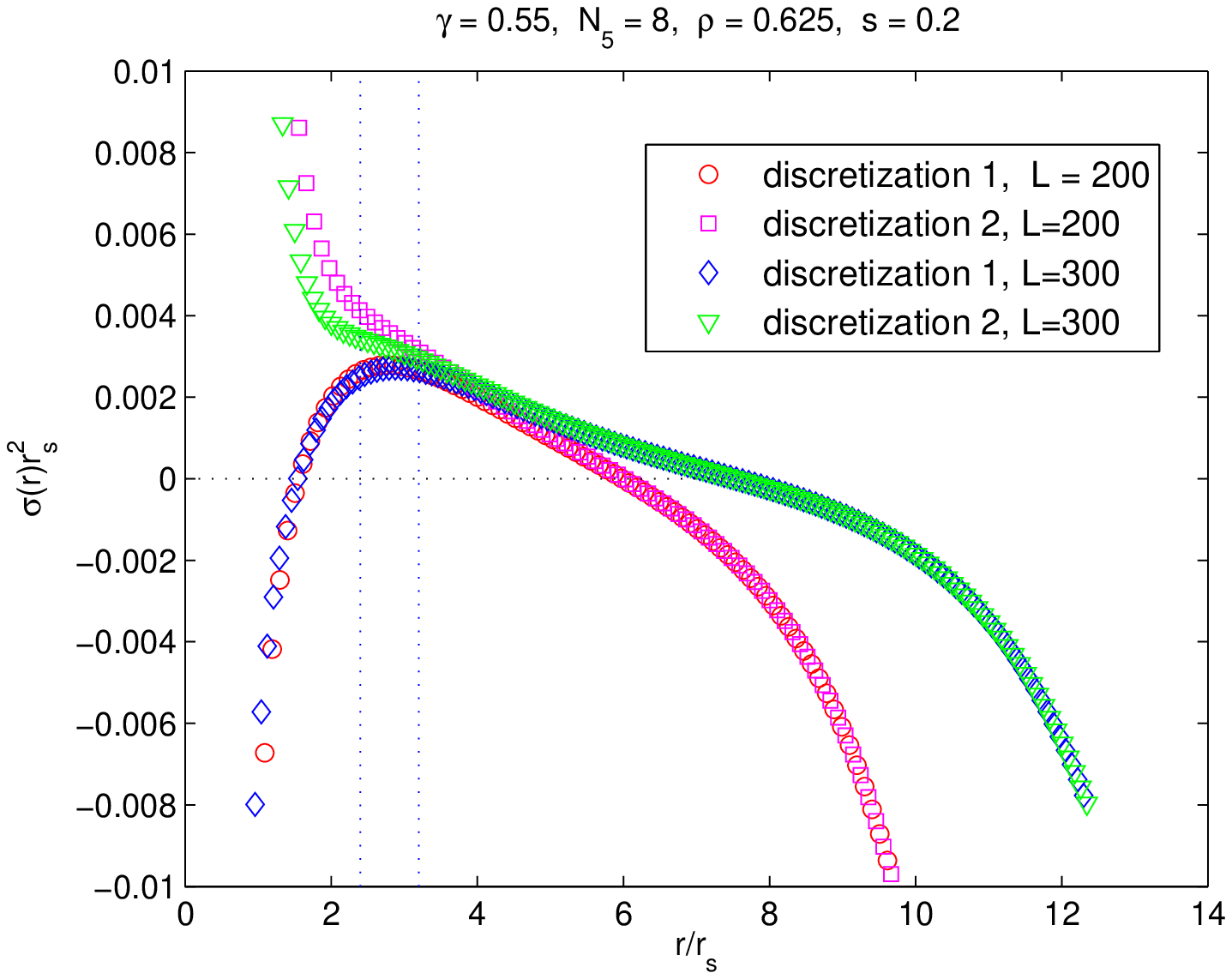}} \ \
  \resizebox{7cm}{!}{\includegraphics{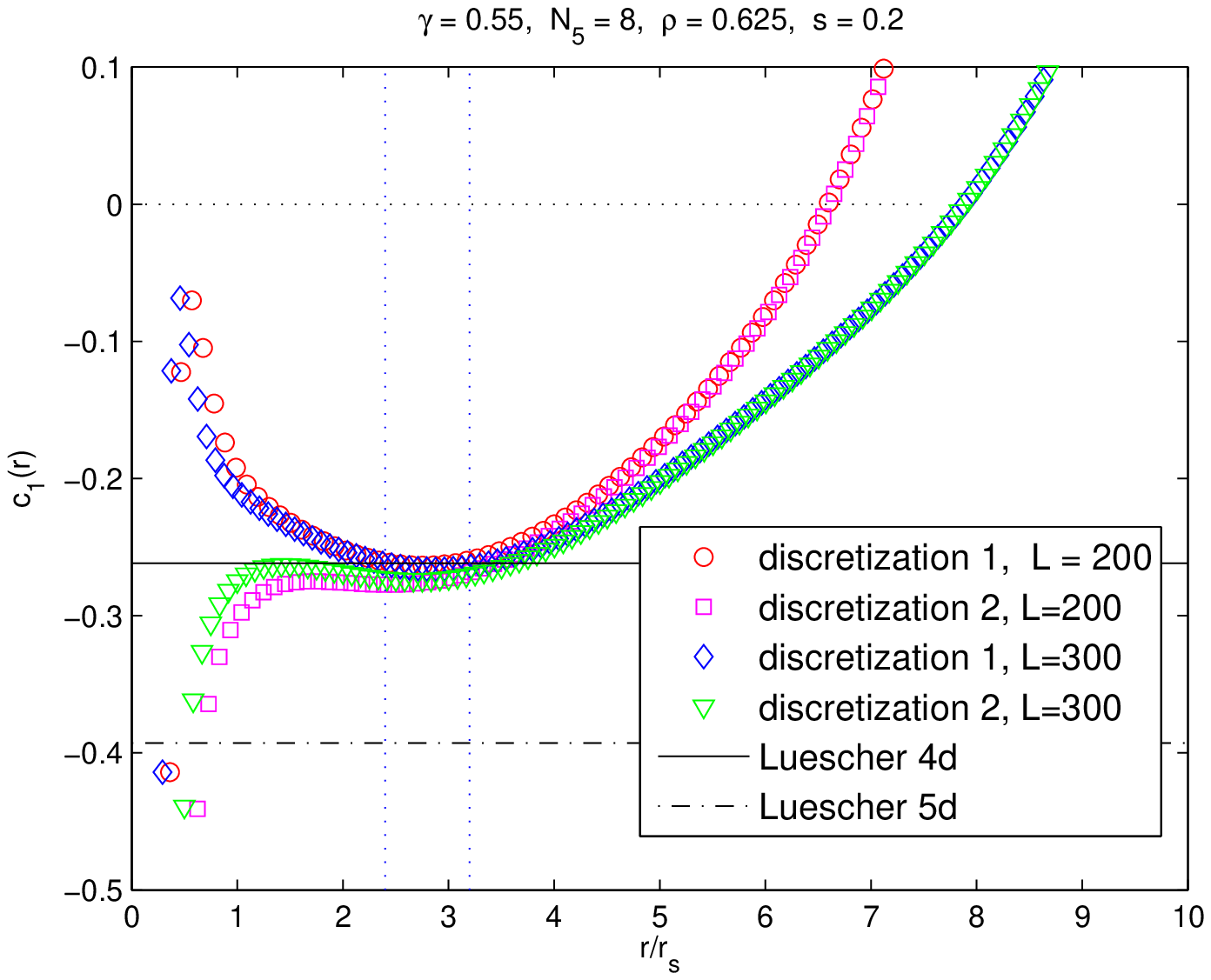}}
  \caption{
The string
    tension $\sigma r_s^2$ and the coefficient $c_1$ of the four-dimensional potential in the middle of the bulk. The coefficients are fitted
    locally using discretization 1\cite{Irges:2009bi} and discretization 2 \cite{Irges:2010qp}.
} 
\label{f_4pbulk}
\end{figure}

We apply the same analysis which has been applied for the torus case to analyze the potential in the middle of the bulk. We do a local fit to $V_4$ of the form
\bea
V_4(r) = \mu + \sigma r + c_0 \log(r) + \frac{c_1}{r} + \frac{c_2}{r^2}\,. 
\eea
We find simultaneous plateaux for the fit coefficients in the region $2.4<r/r_s <3.2 $ and compare two different discretization formulae for the coefficients \cite{Irges:2009bi,Irges:2010qp}. The left plot of \fig{f_4pbulk} is the string tension $\sigma$. The string tension has a positive value. It means there is confinement. The right plot of \fig{f_4pbulk} is the coefficient $c_1$. It shows that the coefficient $c_1$ is consistent with the universal value of the L\"uscher term $-(d-2)\pi/24$ \cite{Luscher:1980fr,Luscher:1980ac} in $d=4$. It is one evidence of dimensional reduction.

\begin{figure}[h!]\centering
  \resizebox{7cm}{!}{\includegraphics{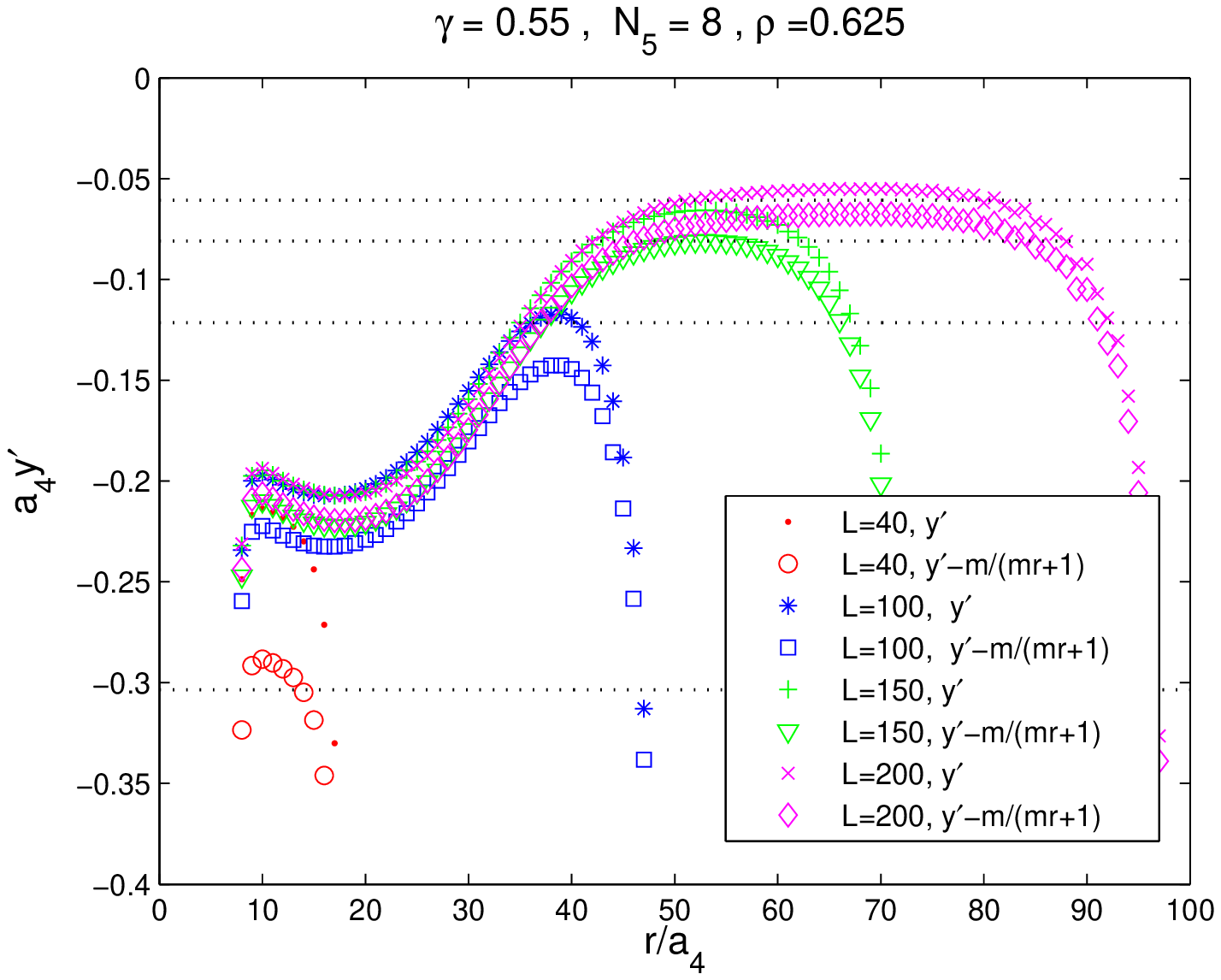}} \ \
  \resizebox{7cm}{!}{\includegraphics{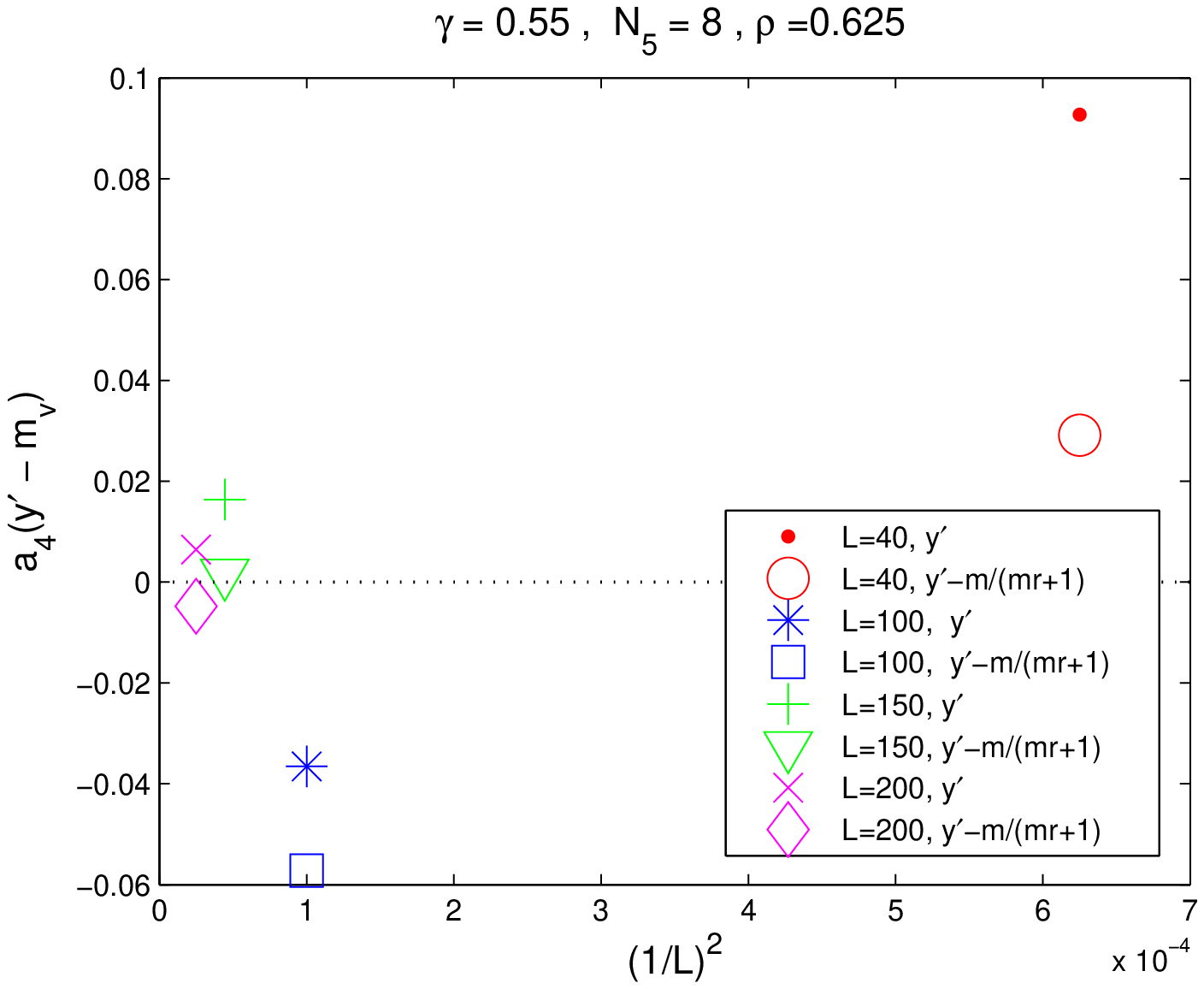}}
  \caption{
    Analysis of the four-dimensional static potential at the boundary. We assume it is a Yukawa potential.
}
\label{f_4pbound}
\end{figure}

We take an ansatz that it is the four-dimensional Yukawa potential with mass $m$ for the potential on the boundary.
\bea
V_4&=&-\alpha  \frac{e^{-mr}}{r} +C,\ \ \alpha > 0 ; \ \ F_4=V_4'=\alpha  \frac{e^{-mr}}{r}(m+\frac{1}{r})  \\
y&=&{\rm log}(r^2F_4)={\rm log}(\alpha)-mr+{\rm log}(mr+1) \\
y'&=&-m+\frac{m}{mr+1}
\eea
Then we numerically compute $y'=(y(r+a_4)-y(r))/a_4$ and see whether we can extract this Yukawa mass. The left plot of \fig{f_4pbound} is $y'$ and  $y' -m/(mr+1)$ for each $L=40,100,150,200$. In this plot we find plateaux which correspond to $-m$ . The dotted lines are the values of the boson mass $m_v$ that we have computed before for the given L. There is a nice agreement between the plateaux values and the dotted lines which represent $c_L/L$ and their difference goes to zero as L goes to infinity (right plot in \fig{f_4pbound}).

\begin{figure}[h!]\centering
  \resizebox{7cm}{!}{\includegraphics{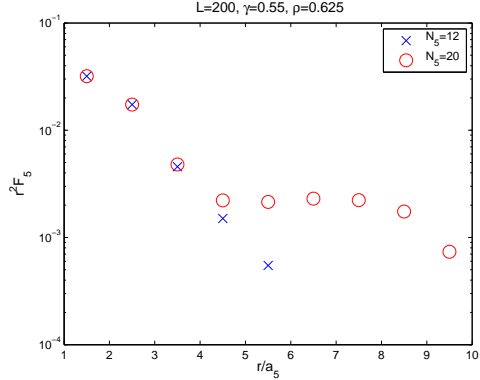}}
  \caption{The static force along fifth dimension.}
  \label{f_5p}
\end{figure}

The plot of \fig{f_5p} is $r^2 F_5$ where $F_5$ is the force along the extra dimension. The plot shows that it is compatible with a five-dimensional Coulomb potential $V_5 \propto 1/r^2 $ for $4 \leq r/a_5 \leq 8$. At small distance, the potential has a different shape. This might be related to the potential along the four-dimensional slices which close to the boundary is different than in the middle of the bulk, as we discussed above. 

\newpage

\section{Conclusion and future works}
In this work using a mean-field calculation we see that there is no spontaneous symmetry breaking on the orbifold because the gauge boson mass goes zero in infinite volume. From the analysis of the four-dimensional static potential in the middle of the bulk for $\gamma=0.55$, we found evidence of confinement and dimensional reduction. 
The four-dimensional static potential at the boundary is compatible with a four-dimensional Yukawa potential whose Yukawa mass corresponds to the vector boson mass in finite volume.

Our next work is to use the twisted orbifold formalism \cite{Scrucca:2003} to study spontaneous symmetry breaking. We are also doing Monte Carlo simulations. Monte Carlo simulation of SU(2) has been done on the orbifold for $\gamma=1$ \cite{Irges:2006qp} and on the torus \cite{Knechtli:2011}. A more realistic model is $SU(3)$ on the orbifold. Due to the boundary condition, the $SU(3)$ gauge symmetry breaks to $SU(2) \times U(1)$ on its boundaries and then spontaneous symmetry breaking might occur. This would correspond to the Standard Model.

\acknowledgments
K. Y is supported by the Marie Curie Initial Training Network STRONGnet. STRONGnet is funded by the European Union under Grant Agreement number 238353 (ITN STRONGnet). We thank A. Kurkela and A. Maas for discussion during the conference.

\end{document}